\documentclass[10pt]{scrartcl}

\usepackage{amsmath, amssymb, amsthm, color, url, fullpage, supertabular}
\usepackage[colorlinks, linkcolor=black, urlcolor=blue, citecolor=black]{hyperref}
\usepackage[neverdecrease]{paralist}

\usepackage{
	natbib,
	algorithm,
	algorithmic
}

\setcitestyle{round}
\newcommand*{\arXiv}[1]{\bgroup\color{blue}\href{http://arxiv.org/abs/#1}{arXiv:#1}\egroup}
\newcommand*{\doi}[1]{\bgroup\color{blue}\href{http://dx.doi.org/#1}{doi:#1}\egroup}
\newcommand*{\email}[1]{\bgroup\color{blue}\href{mailto:#1}{#1}\egroup}
\renewcommand*{\url}[1]{\bgroup\color{blue}\href{#1}{#1}\egroup}

\newcommand{\inactvar}{z}
\newcommand{\actvar}{y}
\newcommand{\fullvar}{x}
\newcommand{\iw}{w} 

\newcommand{\actbas}{B_\textup{a}}
\newcommand{\inactbas}{B_\textup{i}}

\newcommand{\todo}[1]{\bgroup\color{red}\bfseries#1\egroup}

\newcommand*{\ppara}[1]{\noindent\textbf{\textsf{#1}}\,\,}

\usepackage{mathtools}
\newcommand*{\defeq}{\coloneqq}

\newcommand*{\Expect}{\mathbb{E}}

\newcommand*{\quark}{\setbox0\hbox{$x$}\hbox to\wd0{\hss$\cdot$\hss}}
\newcommand*{\rd}{\mathrm{d}}
\newcommand*{\Reals}{\mathbb{R}}
\newcommand*{\targd}{\rho}
\newcommand*{\targfunc}{f}
\newcommand*{\prid}{\rho_{\textup{p}}}
\newcommand*{\liked}{\rho_{\textup{l}}}
\newcommand*{\etargd}{\widehat{\targd}}
\newcommand*{\impd}{q_{\inactvar}}
\newcommand*{\propd}{q_{\actvar}}

\newcommand*{\absval}[1]{\bigl\vert #1 \bigr\vert}

\theoremstyle{definition}

\numberwithin{equation}{section}
\numberwithin{figure}{section}

\begin{document}

\title{Exact active subspace Metropolis--Hastings, with applications to the Lorenz-96 system}

\author{%
	I.\ Schuster\footnote{Institute of Mathematics, Free University of Berlin, and Zuse Institute Berlin, Takustrasse 7, 14195 Berlin, Germany, \email{schuster@zib.de}} %
	\and
	P.\ G.\ Constantine\footnote{Computer Science, University of Colorado Boulder, Boulder, CO 80309, United States of America, \email{paul.constantine@colorado.edu}} %
	\and %
	T.\ J.\ Sullivan\footnote{Institute of Mathematics, Free University of Berlin, and Zuse Institute Berlin, Takustrasse 7, 14195 Berlin, Germany, \email{sullivan@zib.de}} %
}

\date{\today}

\maketitle

\begin{abstract}
	\ppara{Abstract:}
	We consider the application of active subspaces to inform a Metropolis--Hastings algorithm, thereby aggressively reducing the computational dimension of the sampling problem.
	We show that the original formulation, as proposed by Constantine, Kent, and Bui-Thanh (\emph{SIAM J.\ Sci.\ Comput.}, 38(5):A2779--A2805, 2016), possesses asymptotic bias.
	Using pseudo-marginal arguments, we develop an asymptotically unbiased variant.
	Our algorithm is applied to a synthetic multimodal target distribution as well as a Bayesian formulation of a parameter inference problem for a Lorenz-96 system.
	
	\smallskip
	
	\ppara{Keywords:}
	Markov chain Monte Carlo,
	pseudo-marginal,
	active subspace,
	dimension reduction,
	Bayesian parameter inference.
	
	\smallskip
	
	\ppara{2010 Mathematics Subject Classification:}
	65C40, 
	65C05, 
	62H25. 
\end{abstract}

\section{Introduction}
\label{sec:introduction}

Effective sampling of probability density functions on high-dimensional spaces is widely acknowledged as a challenging regime for Markov chain Monte Carlo (MCMC) techniques.
The poor performance of many MCMC schemes in this setting can be attributed to the high dimension leading to both high variance and poor mixing properties, and it is therefore naturally appealing to attempt to improve the convergence properties of MCMC sampling by aggressively reducing the computational dimension to the greatest extent possible while still preserving statistically desirable properties.

Active subspaces \citep{Constantine:2015} are a dimensionality reduction scheme originating in applied mathematics and have recently been applied to Metropolis--Hastings algorithms \citep[ASMH, ][]{ConstantineKentBuiThanh:2016}.
In its original formulation, ASMH was proved to sample from a density $\targd_\epsilon$ that is close to a desired target density $\targd$ in the Hellinger distance.
In this work, we use results from the theory of pseudo-marginal Metropolis--Hastings \citep{Andrieu:2009} to 
\begin{itemize}
	\item establish that indeed ASMH in its original formulation results in a biased sample;
	\item derive a variant that samples \emph{exactly} (in stationarity) from the target;
	\item make exact ASMH applicable when prior and likelihood violate the Gaussianity assumption.
\end{itemize}
When used in a pseudo-marginal setting, active subspaces play the role of a preprocessing technique for adapting the pseudo-marginal sampler to the structure of the posterior.
After considering a synthetic bimodal Gaussian mixture model as a test case, we apply our technique to a Bayesian formulation of a parameter inference problem in the Lorenz-96 model.


Our motivation for the sampling of probability density functions on high-dimensional spaces stems from a desire to solve Bayesian inverse problems.
That is, we have a density $\targd(\fullvar)$ that is given only proportionally by applying Bayes' theorem to a specified probability model and observed data that is assumed to be generated by the model.
We assume the difficulty in this problem is aggravated by the fact that $\fullvar \in \Reals^m$ where $m$ is large.

Indeed, MCMC sampling and its many variants are the computational workhorses of Bayesian inversion, but many problems of practical interest involve high- or even infinite-dimensional $\fullvar$, which is exactly the regime in which many MCMC schemes experience a dramatic deterioration in mixing or convergence properties or increase in autocorrelation and computational cost.
Such problems can be attributed to multimodal and anisotropic structure, and the ease with which a high-dimensional random variable can have high variance:
good mixing of the MCMC chain between well-separated modes is hard to ensure without a priori knowledge;
some adaptation of isotropic proposals to anisotropic structure is essential for acceptable acceptance rates \citep{HaarioSaksmanTamminen:2001, GirolamiCalderhead:2011};
and variance controls the quality of MCMC estimates of integrals against $\targd$.

We propose to use active subspaces to deal with the problem of high dimension \citep{ConstantineKentBuiThanh:2016}.
Concretely, we seek a reparametrization of the domain $\Reals^m$ of $\targd$ into $\actvar \in \Reals^{n}, \inactvar \in \Reals^{m-n}$, such that $\actbas \actvar + \inactbas \inactvar = \fullvar \in \Reals^m$, $n \ll m$ and the marginal density $\targd(\actvar)$ captures most of the variability of $\targd(\fullvar)$.
Here $[\actbas, \inactbas]$ is an orthonormal basis of $\Reals^m$ and usually obtained by eigendecomposing a matrix informed by $\targd(\fullvar)$;
further details are given in Section~\ref{sec:active_subspaces}.
The $\actvar$ are called the \emph{active variables} and $\inactvar$ the \emph{inactive variables}.
When considering integrals of the form $\int_{\Reals^m} \targfunc(\fullvar) \targd(\fullvar) \, \mathrm{d} \fullvar$ that can not be solved analytically, we resort to Monte Carlo techniques.
In this paper we propose to integrate over $\actvar$ with a Metropolis--Hastings-based MCMC algorithm, integrating out $\inactvar$ with a nested numerical integration algorithm.
This is the general method already proposed by \cite{ConstantineKentBuiThanh:2016}.

However, departing from previous work on active subspace Metropolis--Hastings (ASMH), we develop a variant that both takes advantage of active subspaces and at the same time is asymptotically exact.
Not introducing bias through the active subspace approximation while still taking advantage of the induced dimension reduction is of course preferable, especially as no additional computational cost is involved.
To achieve this, we will make use of pseudo-marginal ideas developed by the Bayesian computation community \citep{Beaumont:2003,Andrieu:2009}.
\section{Active subspaces and their use for MCMC}
\label{sec:active_subspaces}

For the purpose of this paper, we will consider a notion of active subspaces tailored to the problem of sampling.
Assume that we are interested in the integral of some function $\targfunc \colon \Reals^m \to \Reals$ with respect to the target density $\targd \colon \Reals^m \to \Reals_+$, i.e.\ $\int_{\Reals^m} \targfunc(\fullvar) \targd(\fullvar) \, \mathrm{d}\fullvar$.
We further assume that this integral exists and is finite.
We will construct matrices $\actbas \in \Reals^{n \times m}, \inactbas \in \Reals^{(m-n) \times m}$ such that
$\begin{bmatrix}
\actbas^\top
\inactbas^\top
\end{bmatrix}^\top$
is an orthonormal basis of $\Reals^m$ and such that either $\targd$ or $\absval{\targfunc} \targd$ vary mostly in the subspace spanned by $\actbas$, while varying much less in the space spanned by $\inactbas$
Hence we call $\actbas$ the basis of the \emph{active subpace}, and $\inactbas$ the basis of the \emph{inactive subspace}.

\subsection{Constructing an active subspace}

There are several methods of constructing an active subspace.
We will go into detail about three of them, namely gradient covariance, posterior covariance and the linear regression method.

\subsubsection{Covariance matrix of log-likelihood gradient}
\label{ssec:Active subspace ll gradient}

The method suggested by \citet{ConstantineKentBuiThanh:2016} for use with Metropolis--Hastings is to estimate the expected outer product of the log-likelihood gradient $\nabla \log(\liked(x))$ with respect to the prior $\prid$, resulting in the estimate
\[
	\widehat{\Sigma} \defeq \frac{1}{N} \sum_{i=1}^{N} \nabla \log(\liked(X_i)) \otimes \nabla \log(\liked(X_i)) \approx \int \nabla \log(\liked(x)) \otimes \nabla \log(\liked(x)) \prid(x) \, \mathrm{d}x ,
\]
which can be acquired by ordinary Monte Carlo through sampling from the prior, $X_i \sim \prid$ for all $i$.
Now one can eigendecompose $\widehat{\Sigma}$ and identify a spectral gap, i.e.\ a group of dominant eigenvalues well separated from the remainder.
The eigenvectors with large eigenvalues are collected into the matrix $\actbas$ and are the basis of the active subspace.
The rest of the eigenvalues are collected in matrix $\inactbas$, building the basis of the inactive subspace.

\subsubsection{Posterior covariance}
\label{ssec:Active subspace posterior cov}

A method to construct an active subspace that can be used even when the gradient of the log-likelihood $\nabla \log(\liked(x))$ is unavailable or costly to compute involves estimating the covariance matrix of the posterior distribution $\int (x-\mu) \otimes (x-\mu) \, \mathrm{d}\targd(x)$, where $\mu \defeq \int x \, \mathrm{d}\targd(x)$.
Again, this can be estimated by using samples from the prior, $X_i \sim \prid$ for all $i$, using the importance sampling trick:
\[
	\widehat{\mu} \defeq \frac{1}{\sum_{j=1}^{N} \iw_i} \sum_{i=1}^{N} \iw_i X_i ,
\]
where $\iw_i \defeq \frac{\targd(X_i)}{\prid(X_i)}$, and
\[
	\widehat{\Sigma} \defeq \frac{1}{\sum_{j=1}^{N} \iw_i} \sum_{i=1}^{N} \iw_i (X_i - \widehat{\mu}) \otimes (X_i - \widehat{\mu}) .
\]

Again we can eigendecompose $\widehat{\Sigma}$, identify a spectral gap and build the active subspace basis $\actbas$ using the eigenvectors with large eigenvalues, and the inactive subspace basis $\inactbas$ using the eigenvectors with small eigenvalue.

The construction of Section~\ref{ssec:Active subspace ll gradient} together with the Gaussian prior assumption was convenient for the original ASMH algorithm \citep{ConstantineKentBuiThanh:2016}.
However, without increasing computational load, our pseudo-marginal approach allows for more freedom in constructing an active subspace for the purpose of using it in Metropolis--Hastings algorithms.
The advantage of using the posterior covariance is that it is more closely related to the actual quantity of interest, namely the posterior distribution.

\subsubsection{Linear regression for a one-dimensional active subspace}
\label{sec:Linear regression for a 1d active subspace}

As an elementary tool for discovering \emph{the} most important direction of variation, \citet{Constantine:2015} suggests using a method inspired by the notion of a \emph{sufficient summary plot} in visual statistics \citep{Cook:1998}.
The idea is to find a linear regression model mapping the $m$-dimensional $\fullvar$ approximately to $\targd(\fullvar)$ (or alternatively the optimal sampling distribution $|\targfunc(\fullvar)|\targd(\fullvar)$).
Having acquired some points $\{X_i\}_{i=1}^N$, either by random sampling or with a deterministic mesh, the idea is to fit the vector $\actbas$ and scalar $b$ in the linear regression model
\[
	\targd(\fullvar) = \actbas \fullvar + b
\]
using $\{(X_i, \targd(X_i))\}_{i=1}^N$ as input and accompanying output pairs.
The fitted $\actbas$ now approximates the direction that explains most of the variation of the target density.
To build the basis for the inactive subspace, we simply complete $\actbas$ to a basis of $\Reals^m$ using the Gram--Schmidt method, resulting in an orthonormal basis.
All of the basis vectors orthogonal to $\actbas$ build a basis for the inactive subspace, $\inactbas$.

\subsection{Usage in Metropolis--Hastings}

Starting from the general notion of active subspaces, the basic proposal of \citet{ConstantineKentBuiThanh:2016} is to run Metropolis--Hastings on the active subspace ($\actvar$ space) instead of the full space ($\fullvar$ space), integrating out the inactive variables $\inactvar$ using a nested numerical integration technique.
For example, \citet{ConstantineKentBuiThanh:2016} used a nested Gauss--Hermite quadrature for the nested integral. 
When considering the move of the Markov chain from a current state $\actvar$ in the active subspace to a proposed state $\actvar'$, \citet{ConstantineKentBuiThanh:2016} compute a new numerical estimate of both marginal densities $\targd(\actvar)$ and $\targd(\actvar')$ for use inside the MH accept/reject step. 
After changing the nested integration technique to importance sampling, the complete original algorithm is given in Algorithm~\ref{algo:Active subspace Metropolis--Hastings} when setting the \emph{originalVersion} variable to \emph{True}.
One of the advantages of using importance sampling is that the evaluation of the target density can be fully parallelized.
For details on the original active subspace Metropolis--Hastings (ASMH) algorithm, see \cite{ConstantineKentBuiThanh:2016}.
\section{Pseudo-marginal MCMC}
\label{sec:pseudomarginal}

Modern pseudo-marginal methods were developed in the context of population genetics by \citet{Beaumont:2003} and their convergence properties have been studied in depth by \citet{Andrieu:2009}.
While pseudo-marginal methods were not developed with reparametrizations of the integration problem in mind, they split the full state space into variables of interest $\actvar$ and nuisance variables $\inactvar$.
This split was originally motivated by inference problems in Bayesian models and the observation that sometimes `nuisance variables' have to be introduced in the probability model solely for technical reasons.

In our setting we can only evaluate the joint density $\targd(\actvar, \inactvar)$ (up to proportionality), but in pseudo-marginal methods we are interested in computing integrals with respect to the marginal density $\targd(\actvar)$ instead.
The idea, very similar to the original ASMH algorithm, is to get a statistical estimate $\widehat{\targd}(\actvar)$  of $\targd(\actvar)$ and use the estimate in place of the actual density.
If $\Expect [ \widehat{\targd}(\actvar)] = \targd(\actvar)$, i.e.\ the estimate is unbiased, then it is possible to construct a Metropolis--Hastings algorithm that will asymptotically produce samples \emph{exactly} following $\widehat{\targd}(\actvar)$.
Because $\widehat{\targd}(\actvar)$ only approximates the marginal density, but the algorithm is asymptotically exact, these types of algorithms have also been called \emph{exact approximate} sampling algorithms.
Concretely, the estimate will be used inside the Metropolis--Hastings acceptance probability:
\[
	\alpha = \min \left( 1, \frac{\widehat{\targd}(\actvar')\propd(\actvar'|\actvar)}{\widehat{\targd}(\actvar)\propd(\actvar'|\actvar)} \right) ,
\]
where $\actvar'$ is the proposed new point in the active subspace and $\propd(\actvar'|\actvar)$ is the density of the proposal.

Whenever $\widehat{\targd}(\actvar')$ is computed only once per proposed point and recycled for later acceptance probability computations (in case the proposal is accepted),
then under the condition that the estimate is unbiased and some further mild conditions,
the resulting sampling algorithm was proved in \cite{Andrieu:2009} to asymptotically sample from the desired marginal density over $\actvar$.
The algorithm class is called grouped independence Metropolis--Hastings \citep[GIMH;][]{Beaumont:2003,Andrieu:2009}.

If, however, we get a new estimate of the marginal density $\targd(\actvar)$ of the current state for each new computation of the Metropolis--Hastings acceptance probability, then this will result in samples that are provably biased.
This class of algorithms is called Monte-Carlo-within-Metropolis \citep[MCwM;][]{Beaumont:2003,Andrieu:2009}.

\section{Exact ASMH}
\label{sec:unbiased_asmh}

The original ASMH paper \citep{ConstantineKentBuiThanh:2016} proved that the invariant distribution of its sampling algorithm, $\targd_\epsilon(\actvar)$, is close in Hellinger distance to the marginal density $\targd(\actvar)$.
In fact, when considering an unbiased randomized method for the nested integration problem (over the inactive variables $\inactvar$), the original ASMH represents a particular instance of a MCwM algorithm and is thus biased\footnote{In fact, the algorithm as stated in \citet{ConstantineKentBuiThanh:2016} might also be read as an asymptotically unbiased GIMH algorithm \citep{Beaumont:2003}. However, the implementation accompanying the paper and personal discussion with the authors clarified that this is not the case.}.
The pseudo-marginal analysis of the original ASMH using a deterministic nested integration technique is slightly less clear, as the question of whether the estimate is unbiased becomes meaningless in that case.

However, it is possible to construct an active-subspace-informed algorithm that will asymptotically sample from $\targd(\actvar)$ exactly, using the pseudo-marginal approach of \citet{Beaumont:2003} and \citet{Andrieu:2009}.
It would be possible to use, for example, Metropolis-within-Gibbs steps \citep{Andrieu:2009} to marginalize over the inactive variables $\inactvar$, which however we leave to future work.
Instead, we will investigate in detail using the importance sampling construction suggested originally by \citet{Beaumont:2003}.
An unbiased estimator $\etargd(\actvar)$ of $\targd(\actvar)$ is constructed using the importance sampling trick $\targd(\actvar) =  \int \frac{\targd(\actbas \actvar + \inactbas \inactvar)}{\impd(\inactvar)} \impd(\inactvar) \textrm{d} \inactvar $.
This enables estimation $\targd(\actvar)$ from a distribution with density $\impd$ which can be chosen to be easy to sample from, under relatively mild assumptions on $\impd$.
One obvious possibility would be to construct $\impd$ utilizing $\inactbas$.
Then the exact ASMH (eASMH) algorithm is given in Algorithm~\ref{algo:Active subspace Metropolis--Hastings} when setting the \emph{originalVersion} variable to \emph{False}.

\begin{algorithm*}[tb]
	\caption{Active subspace Metropolis--Hastings}
	\begin{algorithmic}
		\label{algo:Active subspace Metropolis--Hastings}
		\STATE \textbf{Input:} $\targd(\fullvar)$, initial value $\fullvar_0$, number of samples from active variables $N$, number of pseudo-marginal samples from inactive variabes $M$, boolean flag \emph{originalVersion}
		\STATE \textbf{Output:} linear reparametrization $\actbas, \inactbas$ s.t.\ $\fullvar = \actbas \actvar + \inactbas \inactvar$, samples from $\targd(\fullvar)$
		\STATE
		\STATE $\actbas, \inactbas = \mathrm{ConstructActiveSubspace}(\targd)$
		\STATE solve $\fullvar_0 = \actbas \actvar_0 + \inactbas \inactvar_0$ for $\actvar_0, \inactvar_0$
		\STATE generate $\inactvar_{0,j} \sim \impd$ for $j=1, \dots, M$
		\STATE $\iw_{0,j} = \targd(\actbas \actvar_0 + \inactbas\inactvar_{0,j})/\impd(\inactvar_{0,j})$ for $j=1, \dots, M$
		\STATE $d = \frac{1}{M}\sum_{j=1}^M \iw_{0,j}$
		\FOR{$i$ from $1,\dots, N$}
		\STATE generate $\actvar_i' \sim \propd(\quark|\actvar_{i-1})$
		\STATE generate $\inactvar'_{i,j} \sim \impd$ for $j=1, \dots, M$
		\STATE $\iw'_{i,j} = \targd(\actbas \actvar'_i + \inactbas\inactvar'_{i,j})/\impd(\inactvar'_{i,j})$ for $j=1, \dots, M$
		\STATE $d' = \frac{1}{M}\sum_{j=1}^M \iw'_{i,j}$
		\IF{originalVersion}
			\STATE generate $\inactvar_{i,j} \sim \impd$ for $j=1, \dots, M$
			\STATE $\iw_{i,j} = \targd(\actbas \actvar_i + \inactbas\inactvar_{i,j})/\impd(\inactvar_{i,j})$ for $j=1, \dots, M$
			\STATE $d = \frac{1}{M}\sum_{j=1}^M \iw_{i,j}$
		\ENDIF
		\STATE generate $b \sim \textup{Bernoulli} \bigl( \min \bigl( 1,  d' \propd(\actvar_{i-1}|\actvar_i') \big/ d \, \propd(\actvar_i'|\actvar_{i-1}) \bigr) \bigr)$
		\IF{$b = 1$}
		\STATE Set $d = d', \actvar_i = \actvar_i', \inactvar_{i,j} = \inactvar'_{i,j}, \iw_{i,j} = \iw'_{i,j}$
		\ELSE 
		\STATE Set $\actvar_i = \actvar_{i-1}, \inactvar_{i,j} = \inactvar_{i-1,j}, \iw_{i,j} = \iw_{i-1,j}$
		\ENDIF
		\ENDFOR
		\STATE return $\fullvar$-samples $\actbas \actvar_i + \inactbas\inactvar_{i,j}$ for $i=1,\dots, N$ and $j=1,\dots,M$
	\end{algorithmic}
	
\end{algorithm*}

The main innovation of the new eASMH algorithm, when compared with the original ASMH algorithm (\emph{originalVersion} is \emph{True}), is that no new estimate of the marginal density $\targd(\actvar)$ is acquired at each iteration.
Rather, the previous estimate is recycled, while of course estimating the marginal density of the newly proposed point $\targd(\actvar')$.
Not only does this recycling result in usage of fewer computational resources, but it also leads to the algorithm being unbiased as proved in \cite{Andrieu:2009}.

Another difference, as discussed before, is that in eASMH the estimate $\etargd(\actvar)$ is computed using an unbiased randomized method, namely importance sampling.
While the method for computing the estimate is not strictly fixed in the original ASMH, for the experiments a deterministic integral estimator (Gauss--Hermite) was used.
The decision to use importance sampling in eASMH was taken for theoretical reasons, as in this case well-known pseudo-marginal assumptions and proofs continue to hold.

\section{Experiments}
\label{sec:experiments}

\subsection{Mixture of Gaussians}
\begin{figure}[tb]
	\minipage{0.33\textwidth}
	\includegraphics[width=\linewidth]{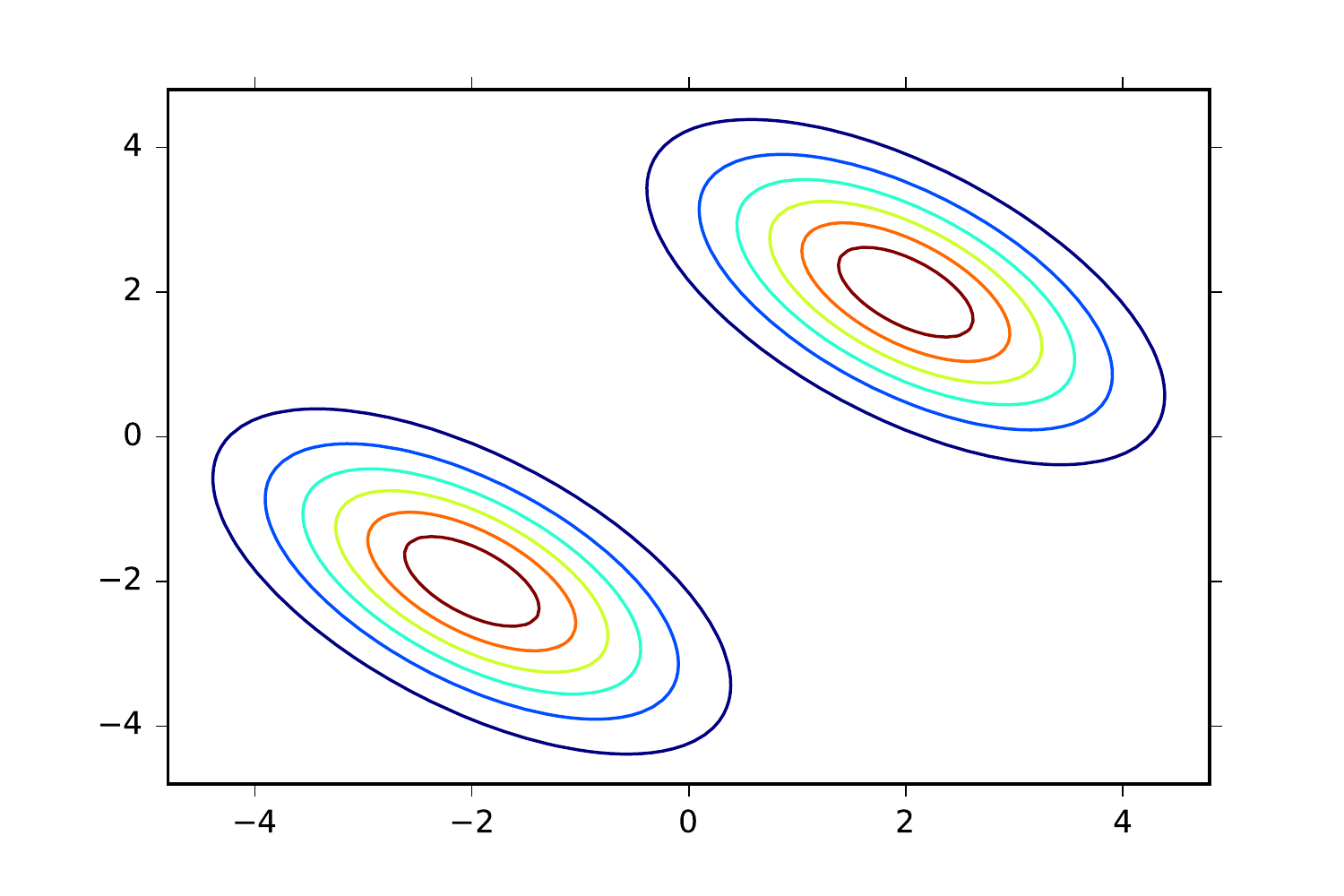}
	
	\endminipage\hfill
	\minipage{0.33\textwidth}
	\includegraphics[width=\linewidth]{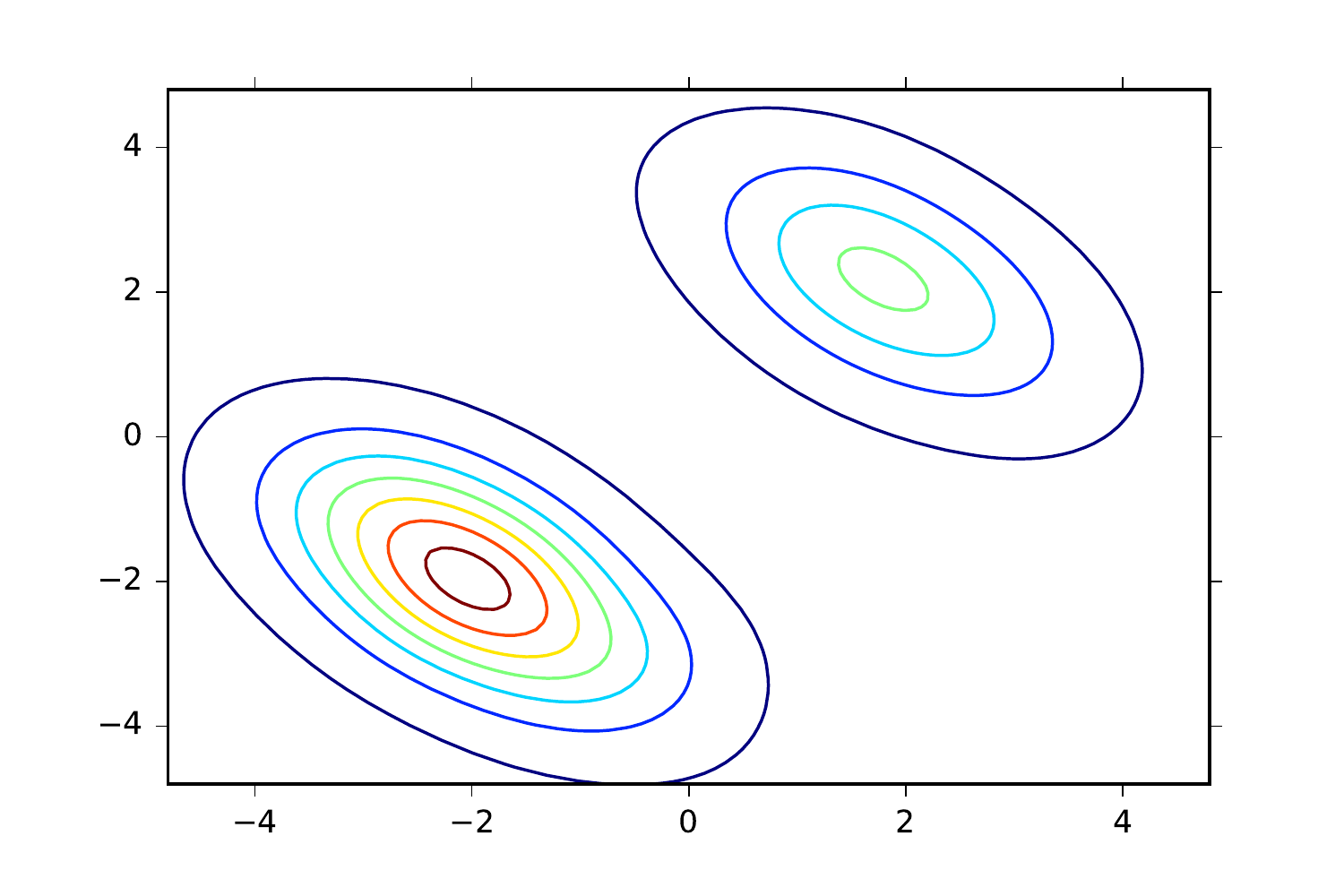}
	
	\endminipage\hfill
	\minipage{0.33\textwidth}%
	\includegraphics[width=\linewidth]{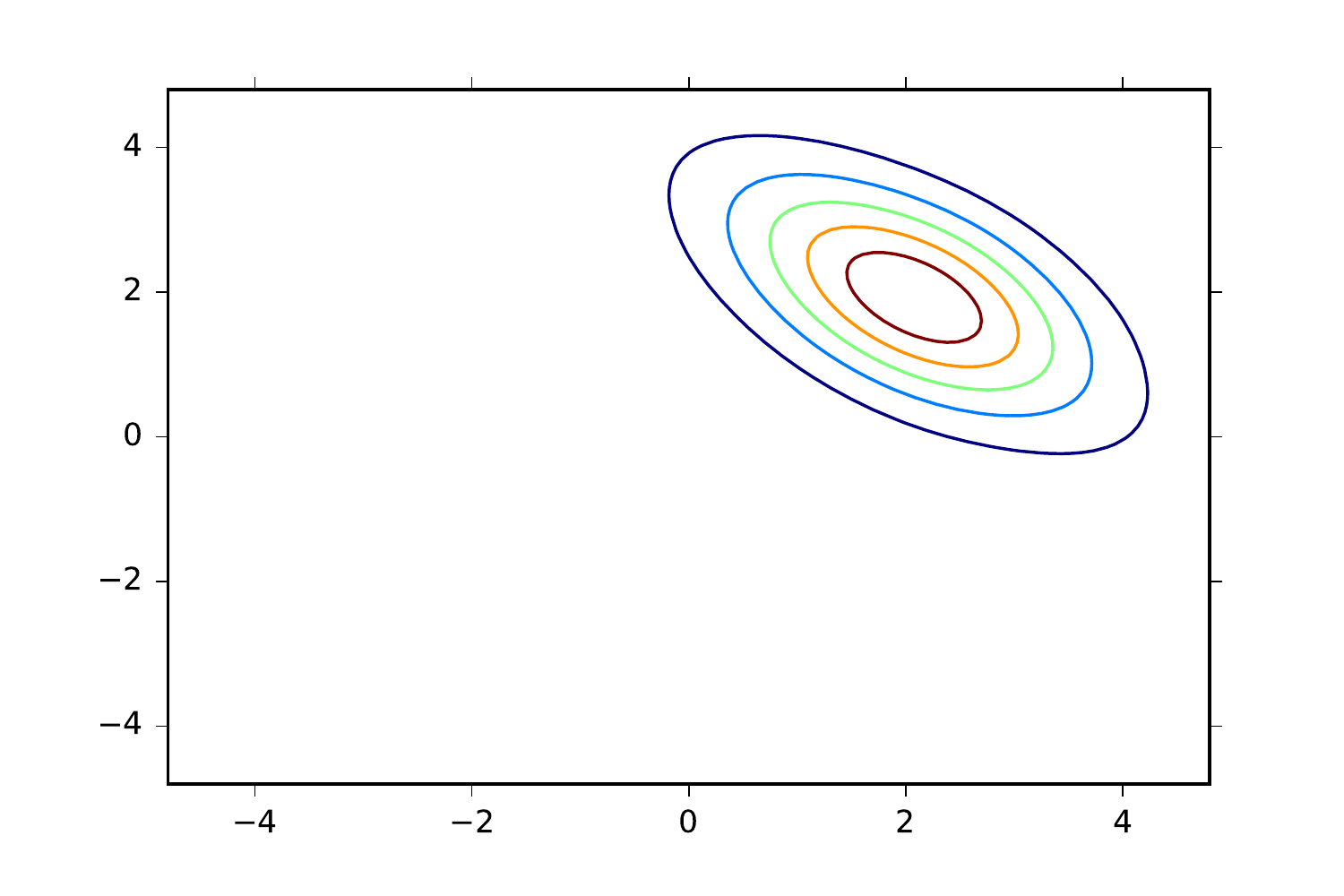}
	\endminipage
	\caption{Convolution of target density with Gaussian density used for the random walk proposal (left), estimate of the same convolution with exact active subspace Metropolis--Hastings (middle) and with vanilla Metropolis--Hastings (right).}
	\label{fig:kde_mog}
\end{figure}

In a first experiment, we used an artificial target density, namely a mixture of two Gaussians in two dimensions, with means and covariances given by
\[
	\mu_1 = - \mu_2 \defeq \begin{bmatrix}
	2\\2
	\end{bmatrix}, \,\, \Sigma_1 = \Sigma_2 \defeq \begin{bmatrix}
	1 & -0.9\\
	-0.9&1
	\end{bmatrix},
\]
and the density itself by
\[
	0.5\, \mathcal{N}(\mu_1, \Sigma_1) + 0.5\, \mathcal{N}(\mu_2, \Sigma_2) .
\]
The resulting bimodal structure is visualized in Figure~\ref{fig:kde_mog} (left).
For the vanilla MH algorithm we used a Gaussian random walk proposal with identity covariance matrix, started at the origin, and ran it for $5500$ iterations, discarding the first $500$ as burn-in.

For the pseudo-marginal active subspace algorithm, we acquired an initial $500$ samples from an isotropic Gaussian centered at the origin with variance $10$.
From these, we built an active subspace approximation using the linear regression method described in Section~\ref{sec:Linear regression for a 1d active subspace}.
The algorithm was then run on the active variable for $500$ iterations with the inactive variable integrated out using $10$ nested samples per proposed point.
This resulted in $5500$ overall target evaluations, just as for the vanilla MH implementation.
The acceptance rate of the vanilla MH algorithm was $32\%$, while the pseudo-marginal active subspace algorithm eASMH accepted $49\%$ of the time, which is close to the optimal acceptance rate \citep{Roberts:2001}.
Thus, while the amounts of overall computation for vanilla MH and eASMH are comparable, the eASMH algorithm demonstrates improved dynamics as quantified by the acceptance rate.

\begin{figure}
	\centering
	\includegraphics[width=0.9\linewidth]{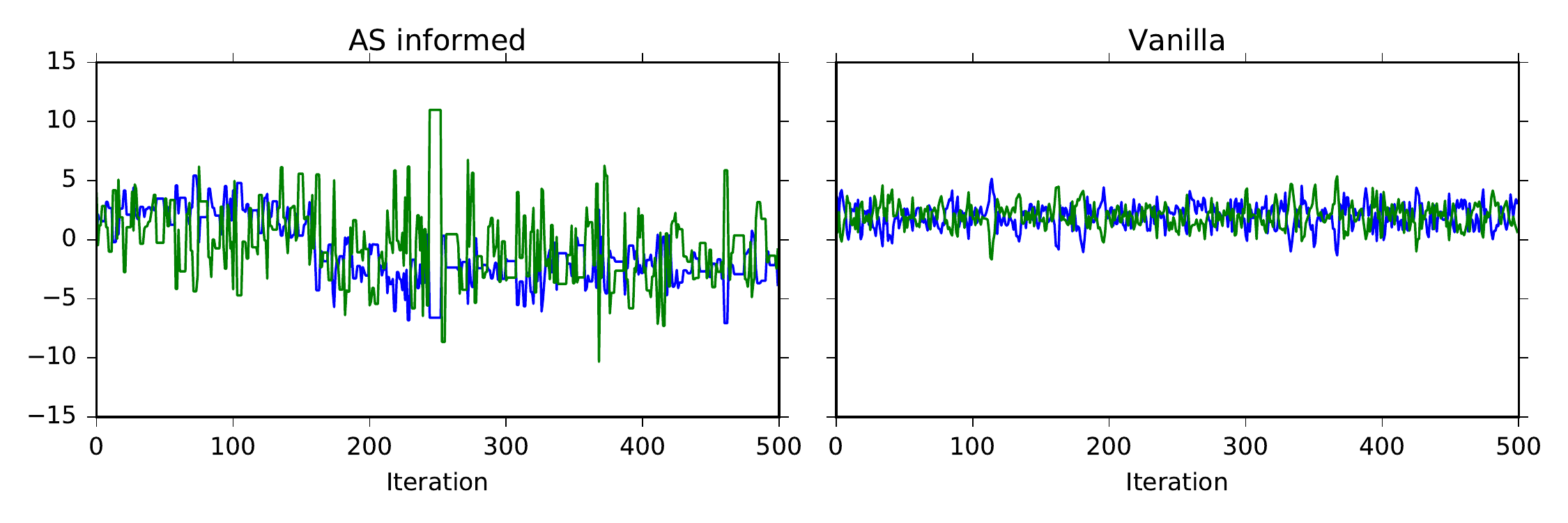}
	\caption{The MH trace plot shows that eASMH explores both modes of the target 2-dimensional Gaussian mixture distribution while vanilla MH becomes stuck in one mode.}
	\label{fig:trace_Gauss}
\end{figure}

\begin{figure}
	\centering
	\includegraphics[width=0.7\linewidth]{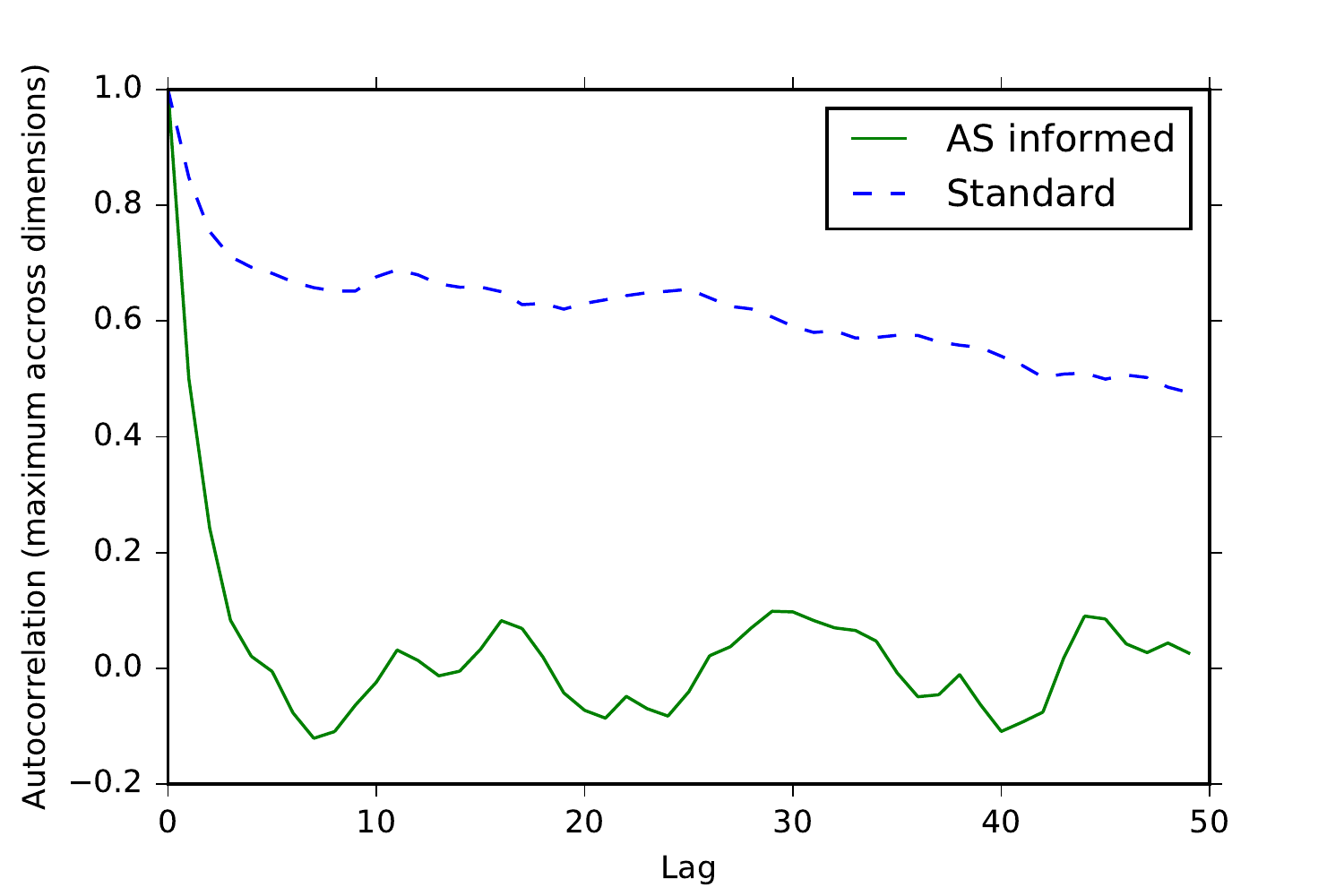}
	\caption{Autocorrelation using eASMH and vanilla MH up to lag 50 for the 2-dimensional Gaussian mixture, computed as the maximum autocorrelation across dimensions, and in each dimension the expected value over the complete chain.}
	\label{fig:Autocorr_Gauss_asvars_10}
\end{figure}

Furthermore, the resulting kernel density estimates are plotted in Figure~\ref{fig:kde_mog} and can be compared with the ground truth target density (which has also convolved with the (Gaussian) density kernel).
Clearly, the eASMH algorithm (middle) is able to recover both modes of the target distribution, which is not the case for the vanilla MH algorithm.
This is also clear from the trace plot in Figure~\ref{fig:trace_Gauss}, which shows how the active subspace algorithm explores all relevant regions of the state space, while the vanilla algorithm gets stuck in one mode.
Quantitatively, this results in lower autocorrelation values for the pseudo-marginal active subspace algorithm (Figure~\ref{fig:Autocorr_Gauss_asvars_10}).

\begin{figure}
	\centering
	\includegraphics[width=0.9\linewidth]{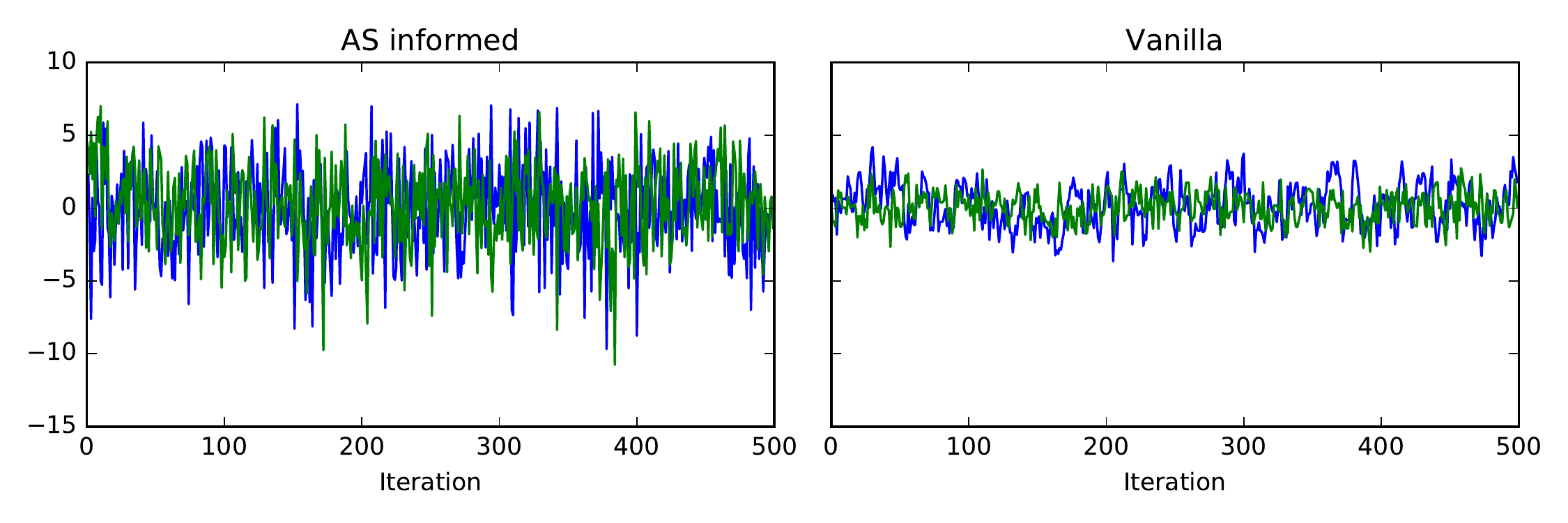}
	\caption{The MH trace plot shows that eASMH explores both modes of the target 10-dimensional Gaussian mixture distribution while vanilla MH becomes stuck in one mode.}
	\label{fig:trace_Gauss_10D}
\end{figure}

\begin{figure}
	\centering
	\includegraphics[width=0.7\linewidth]{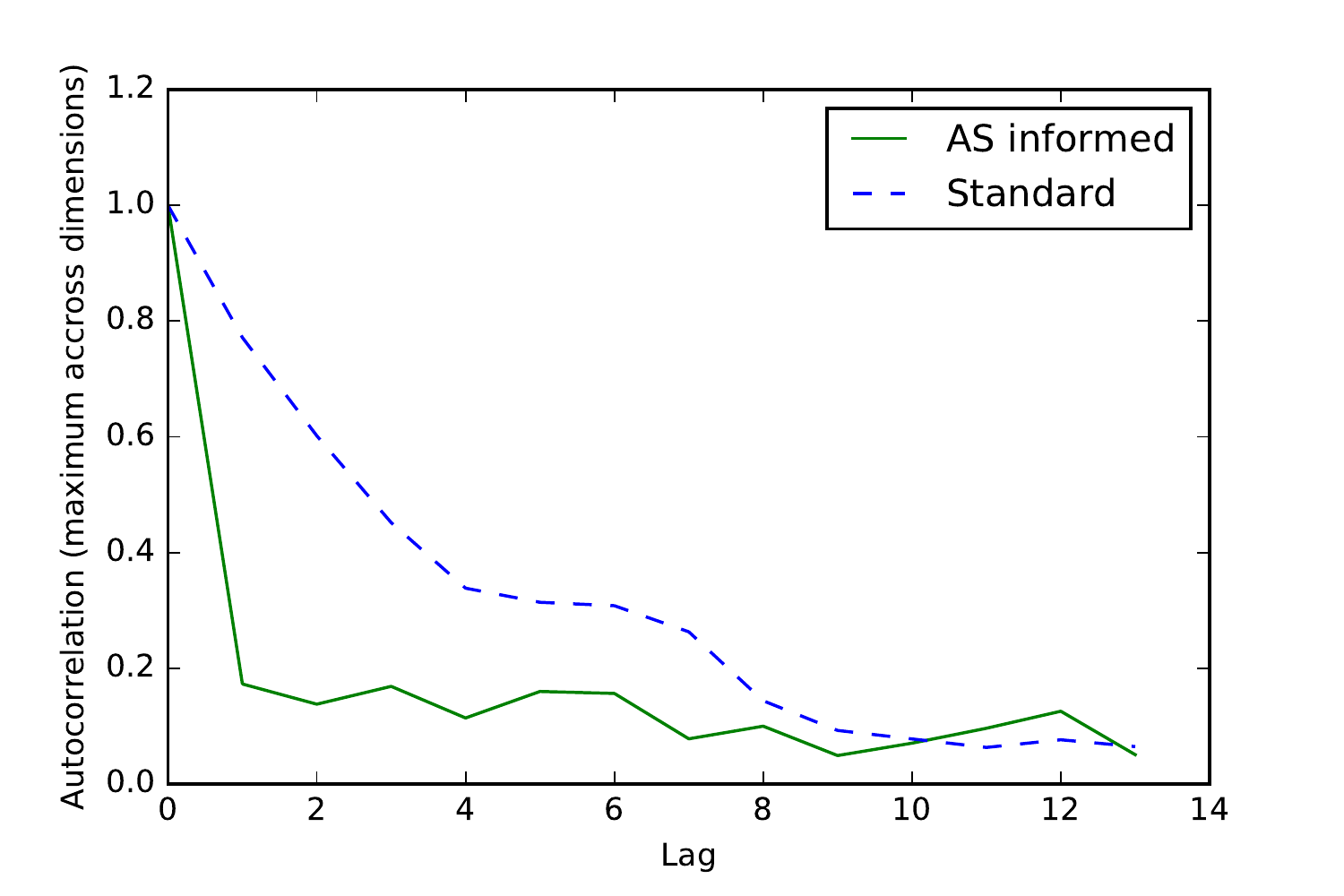}
	\caption{Autocorrelation using eASMH and vanilla MH up to lag 13 for the 10-dimensional Gaussian mixture, computed as the maximum autocorrelation across dimensions, and in each dimension the expected value over the complete chain.}
	\label{fig:Autocorr_Gauss_10D}
\end{figure}

We ran the same algorithm for a $10$-dimensional Gaussian mixture of two components with means coinciding except for the first dimension, which is $2$ or $-2$.
In this case we also the trace plot clearly shows better mixing for the eASMH algorithm as compared to the vanilla MH algorithm (Figure~\ref{fig:trace_Gauss_10D}).
This is reflected in the autocorrelation, which drops more quickly for eASMH than for vanilla MH (Figure~\ref{fig:Autocorr_Gauss_10D}).

\subsection{Lorenz-96 parameter inference}
The Lorenz-96 model is a dynamical system given by the ODEs
\begin{align*}
	\frac{\rd X_{k}}{\rd t} & = - X_{k - 1} ( X_{k - 2} - X_{k + 1} ) - X_{k} + F - \frac{h c}{b} \sum_{j = J ( k - 1 ) + 1}^{k J} Y_{j} \\
	\frac{\rd Y_{j}}{\rd t} & = - c b Y_{j + 1} ( Y_{j + 2} - Y_{j - 1} ) - c Y_{j} + \frac{h c}{b} X_{[ (j - 1) / J ] + 1}
\end{align*}
where the $X_{k}$ variables are large-amplitude, low-frequency variables
that are each coupled to many small-amplitude, high-frequency $Y_{j}$ variables, and $F$ is a forcing constant.
We picked a system in $36$ dimensions.
We fixed a ground truth initial value and computed a forward solve with a forcing constant of $F = 8$, which is known to induce chaotic behavior.
Data $d$ was generated by computing a forward solve from the ground truth using a BDF method with a timestep of $0.01$ between time $0$ and $10$ and adding independent centred Gaussian noise with variance $0.1$ to the all points generated by the forward solve.
We defined a prior density over all inference variables (initial values and the forcing constant) as an isotropic Gaussian centered at $0$ with variance $4$ in every dimension, resulting in a posterior on a $37$-dimensional space.
Given some point $\fullvar$, its likelihood is given by the density of a Gaussian centered at the forward solve $s(\fullvar)$ with isotropic variance $0.1$, evaluated at $d$.

\begin{figure}
	\centering
	\includegraphics[width=0.7\linewidth]{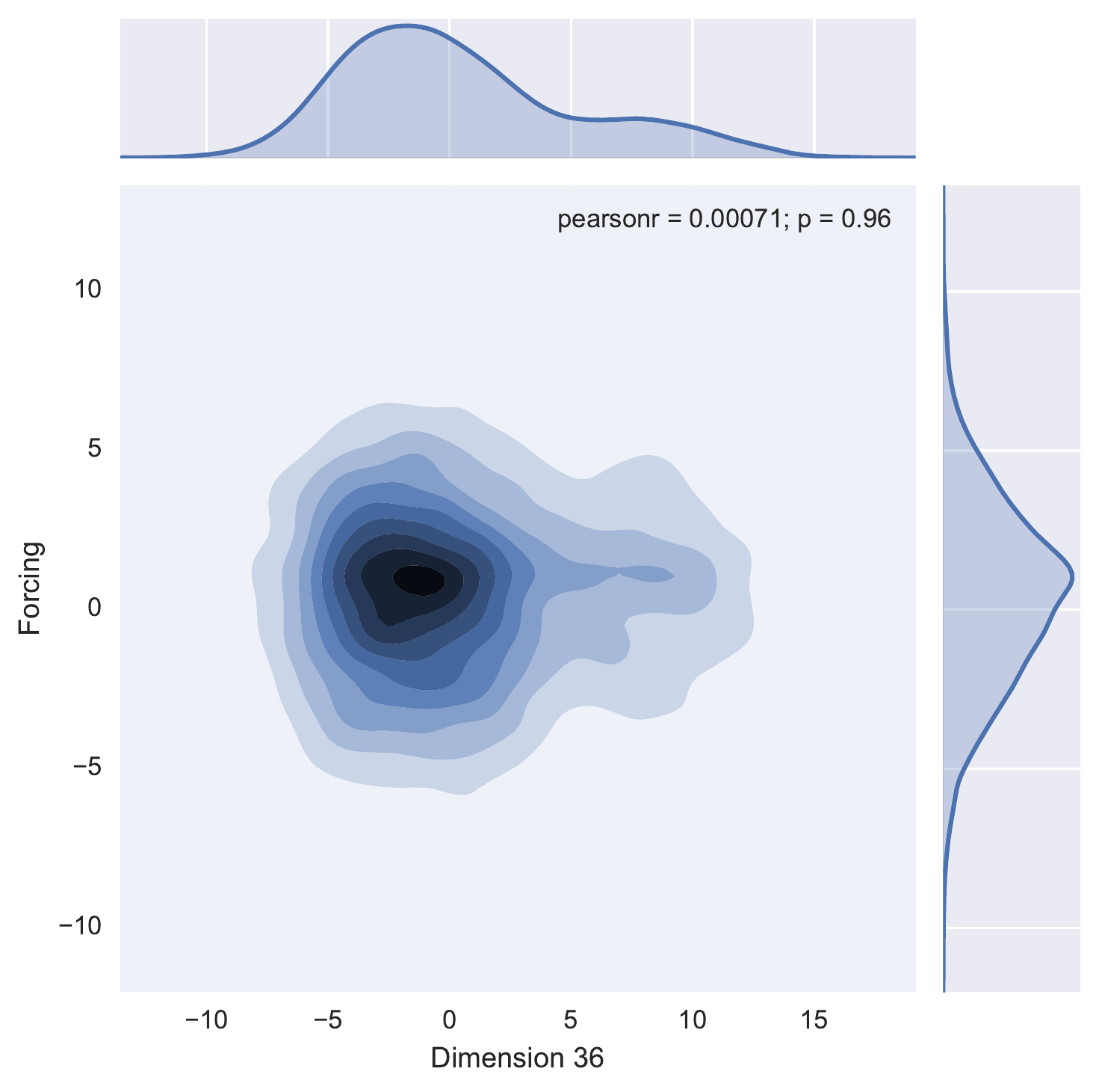}
	\caption{KDE estimate of marginal posterior density of forcing constant and last $x$ component as computed using eASMH samples.}
	\label{fig:Lorenz_marginals_and_joint_kde}
\end{figure}

\begin{figure}
	\centering
	\includegraphics[width=0.7\linewidth]{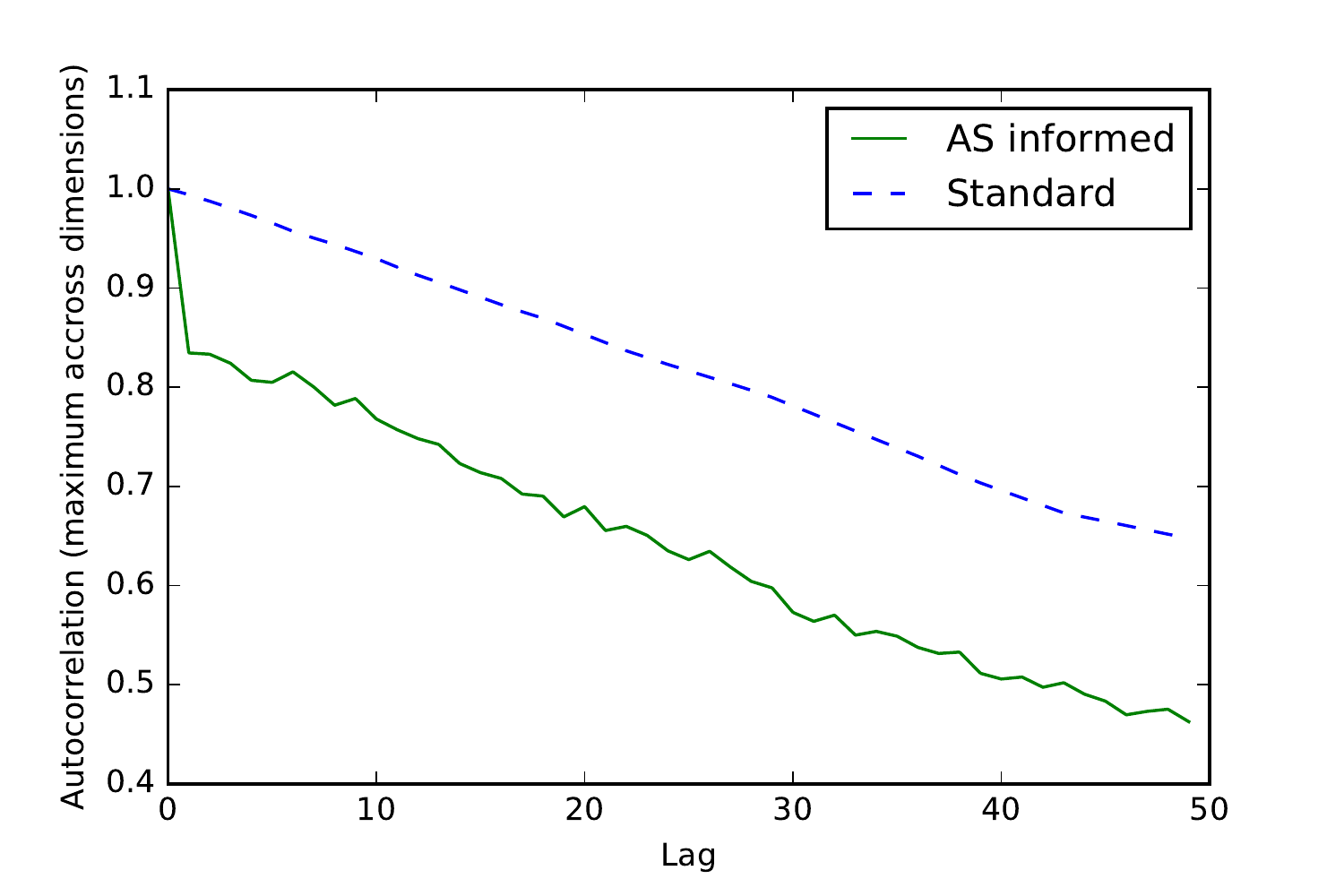}
	\caption{Autocorrelation using eASMH and vanilla MH up to lag $50$ for the Lorenz-96 model, computed as the maximum autocorrelation across dimensions, and in each dimension the expected value over the complete chain.}
	\label{fig:Autocorr_Lorenz96_asvars_8}
\end{figure}

We collected $500$ samples from the prior for constructing the active subspace, using the posterior covariance method outlined in Section~\ref{ssec:Active subspace posterior cov}.
The spectral gap was found between the $4$\textsuperscript{th} and $5$\textsuperscript{th} largest eigenvalues, making the active subspace $4$-dimensional.
The eASMH algorithm was run for $500$ samples in the active subspace, collecting $10$ samples for each to estimate the marginal likelihood.
This makes for $5500$ posterior density evaluations overall, using the same number of forward solves of the ODEs. 
The vanilla MH algorithm was run for $5500$ samples in the full state space, resulting in the same number of posterior density evaluations.
In both cases, a simple Gaussian random walk proposal was used.
The autocorrelation of every $10$\textsuperscript{th} full space sample is given as a function of lag time (i.e.\ number of MH steps) in Figure~\ref{fig:Autocorr_Lorenz96_asvars_8}.
Clearly, eASMH improves upon the short-lag autocorrelation of vanilla MH by a factor of around $0.84$.
A kernel density estimate of the marginal of the forcing constant and the last dimension computed from the eASMH samples is shown in Figure~\ref{fig:Lorenz_marginals_and_joint_kde}.

\section{Conclusions}
\label{sec:conclusions}

In this work, we have used results from the theory of pseudo-marginal Metropolis--Hastings \citep{Andrieu:2009} to conclude that active subspace Metropolis--Hastings in its original formulation results in a biased sample.
We have derived an asymptotically unbiased variant that samples exactly from the target to eliminate this problem.
Exact ASMH (eASMH) is applicable also when prior and likelihood are not Gaussian, which is an assumption made for the original ASMH algorithm.
For eASMH, active subspaces are a preprocessing technique for adapting the pseudo-marginal sampler to the structure of the posterior.
When using importance sampling as the nested Monte Carlo technique, target density evaluations can be computed in parallel, with noticeable speed gains compared to vanilla MH.

Our algorithm strongly decreased autocorrelation compared to vanilla MH when sampling from a mixture of two Gaussians.
For a parameter inference problem in the Lorenz-96 Model, the results also clearly favour exact ASMH over vanilla MH.
One possible area of further improvement is the construction of nonlinear reparametrizations of the full state space, so that nonlinear high posterior density regions can be accurately captured.

\section*{Acknowledgements}
\addcontentsline{toc}{section}{Acknowledgements}

IS and TJS are supported by the German Research Foundation (Deutsche Forschungsgemeinschaft, DFG) through grant CRC 1114 ``Scaling Cascades in Complex Systems'', Project A06, and TJS is further supported by the Free University of Berlin within the Excellence Initiative of the DFG.

\bibliographystyle{abbrvnat}
\addcontentsline{toc}{section}{References}
\bibliography{references.bib}

\begin{thebibliography}{8}
\providecommand{\natexlab}[1]{#1}
\providecommand{\url}[1]{\texttt{#1}}
\expandafter\ifx\csname urlstyle\endcsname\relax
  \providecommand{\doi}[1]{doi: #1}\else
  \providecommand{\doi}{doi: \begingroup \urlstyle{rm}\Url}\fi

\bibitem[Andrieu and Roberts(2009)]{Andrieu:2009}
C.~Andrieu and G.~O. Roberts.
\newblock The pseudo-marginal approach for efficient {M}onte {C}arlo
  computations.
\newblock \emph{Ann. Statist.}, 37\penalty0 (2):\penalty0 697--725, 2009.
\newblock \doi{10.1214/07-AOS574}.

\bibitem[Beaumont(2003)]{Beaumont:2003}
M.~A. Beaumont.
\newblock Estimation of population growth or decline in genetically monitored
  populations.
\newblock \emph{Genetics}, 164\penalty0 (3):\penalty0 1139--1160, 2003.

\bibitem[Constantine(2015)]{Constantine:2015}
P.~G. Constantine.
\newblock \emph{{A}ctive {S}ubspaces: {E}merging {I}deas in {D}imension
  {R}eduction for {P}arameter {S}tudies}, volume~2 of \emph{SIAM Spotlights}.
\newblock Society for Industrial and Applied Mathematics (SIAM), Philadelphia,
  PA, 2015.

\bibitem[Constantine et~al.(2016)Constantine, Kent, and
  Bui-Thanh]{ConstantineKentBuiThanh:2016}
P.~G. Constantine, C.~Kent, and T.~Bui-Thanh.
\newblock Accelerating {M}arkov chain {M}onte {C}arlo with active subspaces.
\newblock \emph{SIAM J. Sci. Comput.}, 38\penalty0 (5):\penalty0 A2779--A2805,
  2016.
\newblock \doi{10.1137/15M1042127}.

\bibitem[Cook(1998)]{Cook:1998}
R.~D. Cook.
\newblock \emph{Regression {G}raphics}.
\newblock Wiley Series in Probability and Statistics: Probability and
  Statistics. John Wiley \& Sons, Inc., New York, 1998.
\newblock \doi{10.1002/9780470316931}.

\bibitem[Girolami and Calderhead(2011)]{GirolamiCalderhead:2011}
M.~Girolami and B.~Calderhead.
\newblock Riemann manifold {L}angevin and {H}amiltonian {M}onte {C}arlo
  methods.
\newblock \emph{J. R. Stat. Soc. Ser. B Stat. Methodol.}, 73\penalty0
  (2):\penalty0 123--214, 2011.
\newblock \doi{10.1111/j.1467-9868.2010.00765.x}.

\bibitem[Haario et~al.(2001)Haario, Saksman, and
  Tamminen]{HaarioSaksmanTamminen:2001}
H.~Haario, E.~Saksman, and J.~Tamminen.
\newblock An adaptive {M}etropolis algorithm.
\newblock \emph{Bernoulli}, 7\penalty0 (2):\penalty0 223--242, 2001.
\newblock \doi{10.2307/3318737}.

\bibitem[Roberts and Rosenthal(2001)]{Roberts:2001}
G.~O. Roberts and J.~S. Rosenthal.
\newblock Optimal scaling for various {M}etropolis--{H}astings algorithms.
\newblock \emph{Statist. Sci.}, 16\penalty0 (4):\penalty0 351--367, 2001.
\newblock \doi{10.1214/ss/1015346320}.

\end{thebibliography}

\end{document}